\documentclass{aa520}
\usepackage{graphicx}
\usepackage{txfonts}
\newcommand{\thestar}{$\varepsilon$\,Indi}
\newcommand{\thetdwarf}{$\varepsilon$\,Indi\,B}
\newcommand{\micron}{\mbox{\,${\mu}$m}}               
\newcommand{\Lsolar}{\mbox{\,$L_{\odot}$\/}}          
\newcommand{\Rsolar}{\mbox{\,$R_{\odot}$\/}}          
\newcommand{\Mjup}{\mbox{\,$M_{\rm Jup}$\/}}          
\newcommand{\magnit}[2]{\mbox{$\mbox{\rm #1}^{\mbox{\rm\tiny m}}%
     \!\!\!.\!\,\, \mbox{\rm #2}$}}                   
\newcommand{\magap}[1]{\mbox{$\mbox{\rm #1}^{\mbox{\rm\tiny m}}$}} 
\newcommand{\oversim}[2]{\lower0.5ex\vbox{\baselineskip=0pt\lineskip=0.2ex
     \ialign{$\mathsurround=0pt #1\hfil##\hfil$\crcr#2\crcr\sim\crcr}}}
\newcommand{\etal}{\mbox{\hbox{et\,al.}}}         
\newcommand{\idest}{\mbox{\hbox{\it i.e.,}}}          
\newcommand{\cf}{\mbox{\hbox{\it cf.}}}               
\begin{document}
\title{\thetdwarf: a new benchmark T dwarf
\thanks{Based on data from the SuperCOSMOS Sky Surveys and
observations collected with the ESO NTT, La Silla, Chile.} 
}

\author{R.-D. Scholz\inst{1} \and
        M. J. McCaughrean\inst{1} \and
        N. Lodieu\inst{1} \and
        B. Kuhlbrodt\inst{2}}

\offprints{R.-D. Scholz; {\tt rdscholz@aip.de}}

\institute{$^1$Astrophysikalisches Institut Potsdam, 
           An der Sternwarte 16, 14482 Potsdam, Germany \\
           $^2$Hamburger Sternwarte, Universit\"at Hamburg, 
           Gojenbergsweg 112, 21029 Hamburg}

\date{Received ...; accepted ...}

\titlerunning{\thetdwarf: a new benchmark T-dwarf}
\authorrunning{Scholz \etal{}}

\abstract{
We have identified a new early T dwarf only 3.6\,pc from the Sun, as a 
common proper motion companion (separation 1459\,AU) to the K5V star 
\thestar{} (HD\,209100). As such, \thetdwarf{} is one of the highest proper 
motion sources outside the solar system ($\sim$\,4.7 arcsec/yr), part of one 
of the twenty nearest stellar systems, and the nearest brown dwarf to the 
Sun. Optical photometry obtained from the SuperCOSMOS Sky Survey was combined 
with approximate infrared photometry from the 2MASS Quicklook survey data 
release, yielding colours for the source typical of early T dwarfs. Follow-up 
infrared spectroscopy using the ESO NTT and SOFI confirmed its spectral 
type to be T2.5$\pm$0.5. With $K_s$=\magnit{11}{2}, \thetdwarf{} is 1.7 
magnitudes brighter than any previously 
known T dwarf and 4 magnitudes brighter than the typical object in its
class, making it highly amenable to detailed study. Also, as a companion to 
a bright nearby star, it has a precisely known distance (3.626\,pc) and 
relatively well-known age (0.8--2\,Gyr), allowing us to estimate its 
luminosity as log\,L/\Lsolar=$-4.67$, its effective temperature as 1260\,K, 
and its mass as $\sim$\,40--60\Mjup{}. \thetdwarf{} represents an important 
addition to the census of the Solar neighbourhood and, equally importantly, 
a new benchmark object in our understanding of substellar objects.
}
\maketitle

\section{Introduction}
Our knowledge of the solar neighbourhood remains remarkably sketchy. Recent 
discoveries of very nearby ($<$10\,pc) early and mid M dwarfs (Scholz, 
Meusinger, \& Jahrei{\ss} \cite{scholz01}; Reid \& Cruz \cite{reid02a}; 
Reid, Kilkenny, \& Cruz \cite{reid02b}; Reyl\'{e} \etal{} \cite{reyle02};
L\'{e}pine, Rich, \& Shara \cite{lepine02a}) and late M dwarfs
(Delfosse \etal{} \cite{delfosse01}; McCaughrean, Scholz, \& Lodieu 
\cite{mccaughrean02}; L\'{e}pine \etal{} \cite{lepine02b}) appear to
vindicate predictions that up to 30\% of all stars within 10\,pc remain 
unknown (Henry \etal{} \cite{henry97}).

The deficiency is even larger at substellar masses. Observational evidence
(Reid \etal{} \cite{reid99}) and theory (Chabrier \cite{chabrier02}) suggest 
a local space density of $\sim$0.1 per cubic parsec for brown dwarfs, \idest{} 
twice that for main sequence stars. However, brown dwarfs are much cooler and 
fainter and thus harder to observe, and as a consequence, only one M9 brown 
dwarf (Tinney \cite{tinney96}, \cite{tinney98}), a handful of L dwarfs, and 
10 T dwarfs (see Burgasser \etal{} \cite{burgasser03} and references therein) 
are known at less than 10\,pc from the Sun, compared to $>$300 stars within 
the same volume.

These sources are important however, as the detailed observation of very 
nearby stars and brown dwarfs is a key starting point for investigations of 
the star formation process, the stellar and substellar luminosity function, 
and the initial mass function. The recently discovered and categorised T 
dwarfs, with temperatures of $\sim$\,1500\,K or less (Burgasser \etal{}
\cite{burgasser02}) are particularly interesting, as they span a range of 
masses from just substellar down close to those of giant planets. Some 32 
spectrally confirmed T dwarfs have been found to date, most of them in 
wide-field infrared (2MASS; Burgasser \etal{} \cite{burgasser02}) and optical 
(SDSS; Geballe \etal{} \cite{geballe02}) surveys. However, as substellar brown 
dwarfs continuously cool and decrease in luminosity throughout their lifetime, 
it is relatively difficult to determine their distances, ages, and masses. 
There are two important exceptions, Gl229\,B and Gl570\,D, both of which are 
companions to nearby stars for which distances and ages are relatively well 
known, and these key template objects help underpin studies of brown dwarf 
evolution, chemistry, and weather.
It is crucial that we expand the sample of T dwarfs with 
well-understood physical properties, and the nearer they are to the Sun,
the better, as they can be studied in more detail. In this paper, we describe 
the discovery of a T dwarf as as a companion to one of the nearest stars in 
the sky, \thestar.

\section{Proper motion data and optical photometry}
There is ample evidence that surveys for high proper motion objects can 
successfully contribute to the completion of the census of the solar 
neighbourhood at the stellar/substellar boundary (Ruiz, Leggett, \& Allard
\cite{ruiz97}; Scholz \etal{} \cite{scholz01}; Reyl\'e \etal{} \cite{reyle02};
Reid \& Cruz \cite{reid02a}; Reid \etal{} \cite{reid02b};
Cruz \& Reid \cite{cruz02}).
We have been using the SuperCOSMOS Sky Survey (hereafter SSS: Hambly \etal{} 
\cite{hambly01a,hambly01b,hambly01c}), a recent scan of the entire southern sky 
made from UKST and ESO Schmidt plates in three passbands ($B_J$, $R$, and 
$I$), to identify new, high proper motion objects, including cool white 
dwarfs, M dwarfs, and L dwarfs (Scholz \etal{} 
\cite{scholz02a}; McCaughrean \etal{} \cite{mccaughrean02}; Lodieu, Scholz,
\& McCaughrean \cite{lodieu02}; Scholz \& Meusinger \cite{scholz02c}).

Scholz \& Meusinger (\cite{scholz02c}) showed that all known L dwarfs within 
20\,pc could be recovered in the SSS data, at least in the $I$ band. In 
addition, it was possible to determine proper motions for them, provided 
multi-epoch SSS measurements were available. We are now searching for further 
high proper motion brown dwarf candidates in the SSS data, applying the 
following selection procedure.

We considered the declination zone 
$-57^{\circ} 30^m < \delta < -37^{\circ} 30^m$, where UKST $B_J$, $R$, and
$I$, and ESO Schmidt $R$ plate data are all available. Starting with the
UKST $I$ data, we selected stars with $I$\,$<$\magap{17} which could not 
be identified with an object on the corresponding UKST $B_J$ and ESO $R$ 
plates within a search radius of 6 arcsec. These latter plates were normally 
taken about 10--15 years earlier than the corresponding UKST $R$ and $I$ 
plates. We then narrowed down the sample to just red objects with UKST 
$R-I>$\,\magnit{2}{2} and, using 1$\times$1 arcmin SSS finding charts, we 
checked to see if the non-identification on the UKST $B_J$ and ESO $R$
plates was due to a measurement error, an extremely red colour, or a large 
proper motion. Objects measured only on the $I$ plate and with no clear 
counterparts in the other passbands were also checked using $I$, $R$, and 
$B_J$ data from overlapping plates. 
Finally, suspected extremely red (large $B_J - R$ or no $B_J$ measurement)
sources with high proper motions derived from linear fits to the multi-epoch
SSS data were examined in the near-infrared using the 2MASS database.

Thus, \thetdwarf{} was initially found as a bright $I$ band source with a
matched object on the UKST $R$ plate, but with no identification on the 
$B_J$ or ESO $R$ plates. Visual inspection of the finding charts revealed
that the UKST $R$ measurement was in fact spurious, the diffraction spike
of a nearby bright star. However, on an overlapping $I$ plate with an epoch
difference of just two years, the source had clearly shifted. The huge 
computed proper motion of 4.7 arcsec/yr led us to check the 2MASS image data,
which revealed a very bright infrared source at the expected position.

\begin{figure}
\centering
\includegraphics[width=3.45in]{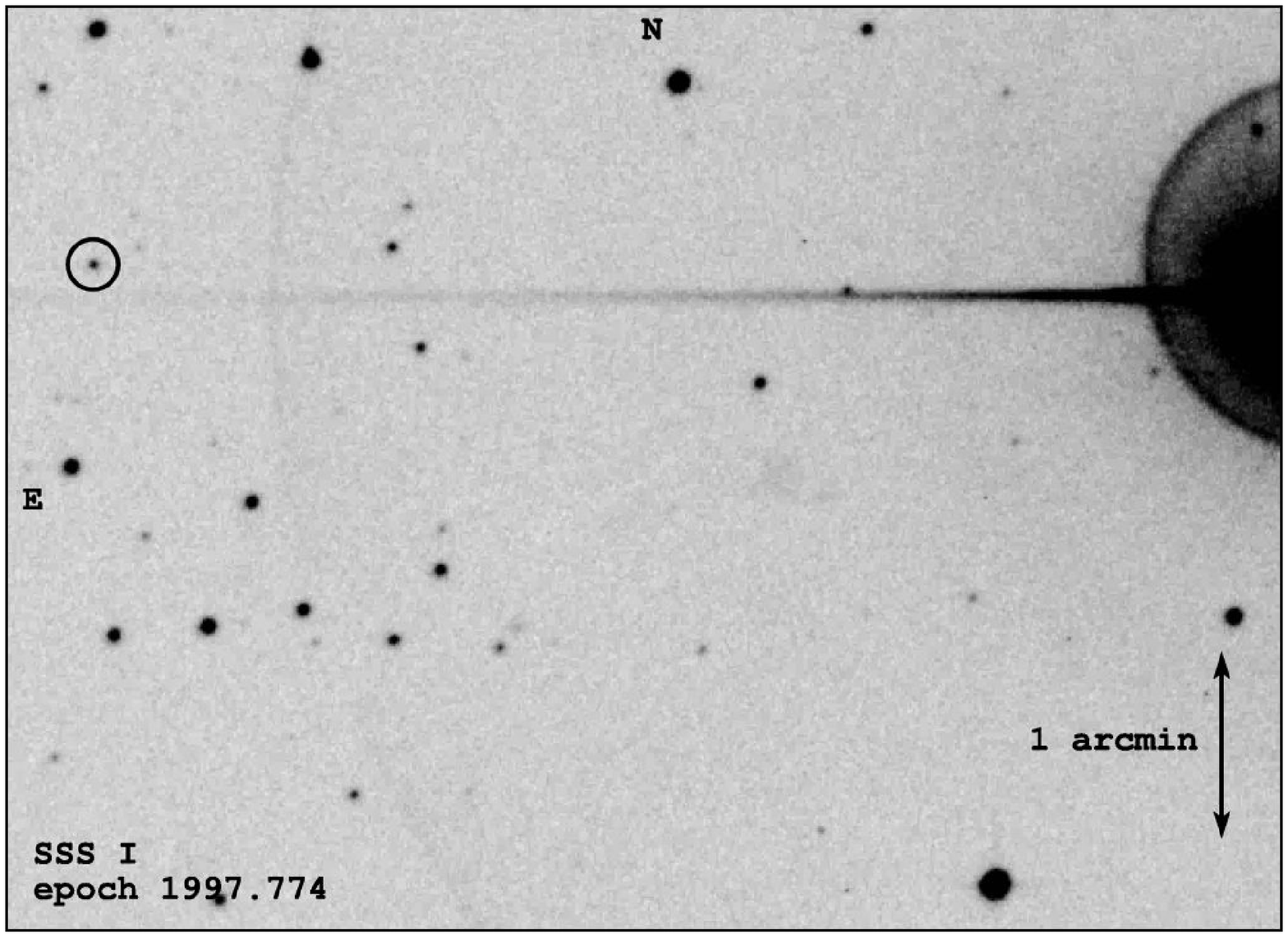}
\vspace{3mm}
\includegraphics[width=3.45in]{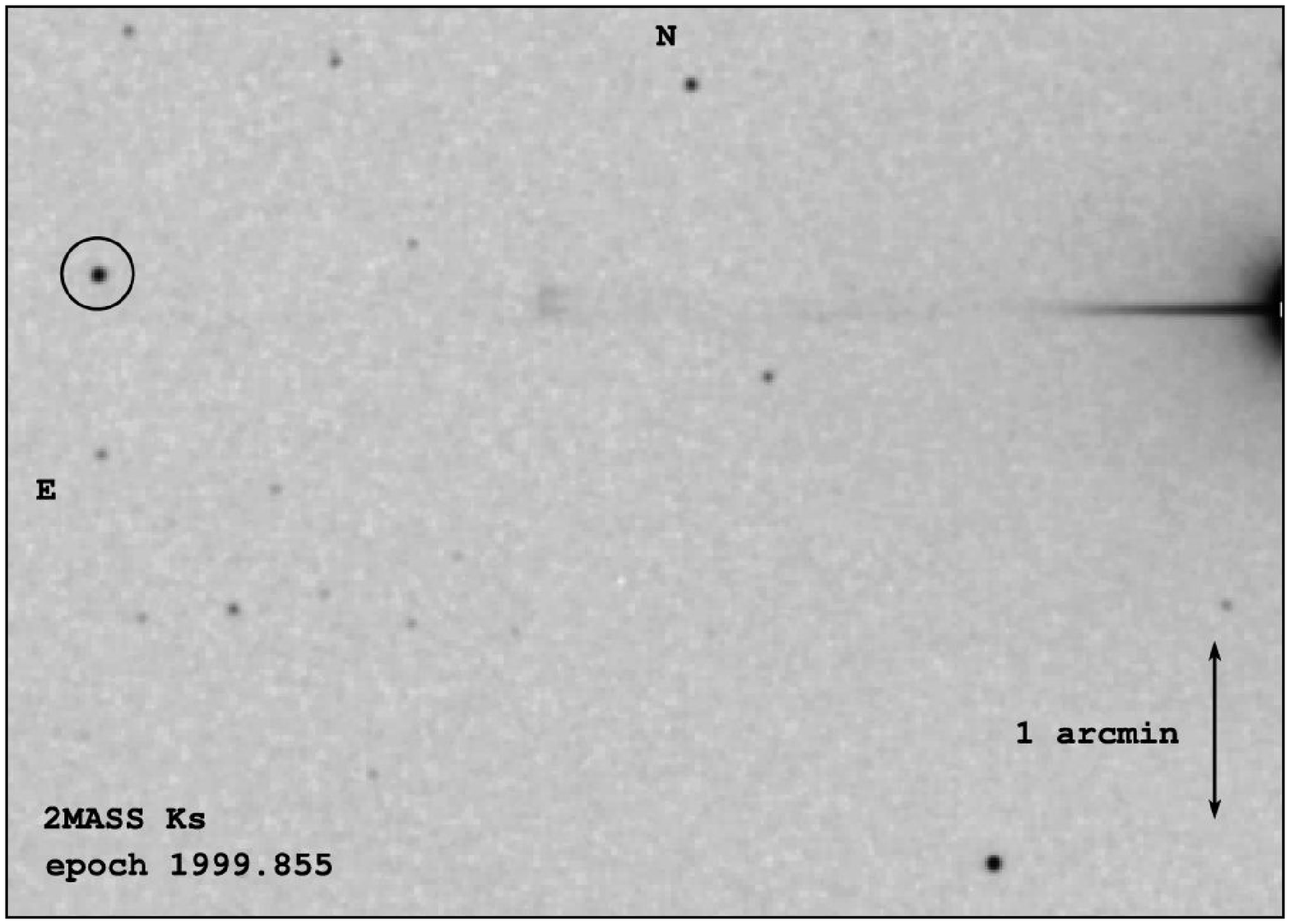}
\caption{Finding chart images for \thestar{} (bright star at far right)
and \thetdwarf{} (circled).
The upper panel is the SSS $I$ plate; the lower is the 2MASS
Quicklook atlas $K_s$ image. The extremely red optical-infrared
colour of \thetdwarf{} is obvious.}
\label{fig:finder}
\end{figure}

Finally, through a catalogue search, we `discovered' \thestar{} (HD\,209100, 
GJ\,845; spectral type K5V) about 7 arcmin away and with exactly the same 
large proper motion (see Figure~\ref{fig:finder}). Thus, it is unequivocal
that these two sources are related, \idest{} that \thetdwarf{} is a 
wide companion to \thestar, and together they constitute one of the twenty
nearest stellar systems in the sky. The primary star has a precise HIPPARCOS 
parallax yielding a distance of 3.626$\pm$0.001\,pc (ESA 1997), although for 
\thetdwarf{} the uncertainty is somewhat larger due to the unknown separation
between the primary and the companion along the line-of-sight: the separation 
in the plane of the sky is 402.3 arcsec or 1459\,AU (epoch 2000.0), 
\idest{} 0.007\,pc.

\begin{figure*}
\centering
\includegraphics[angle=-90,width=5.6in]{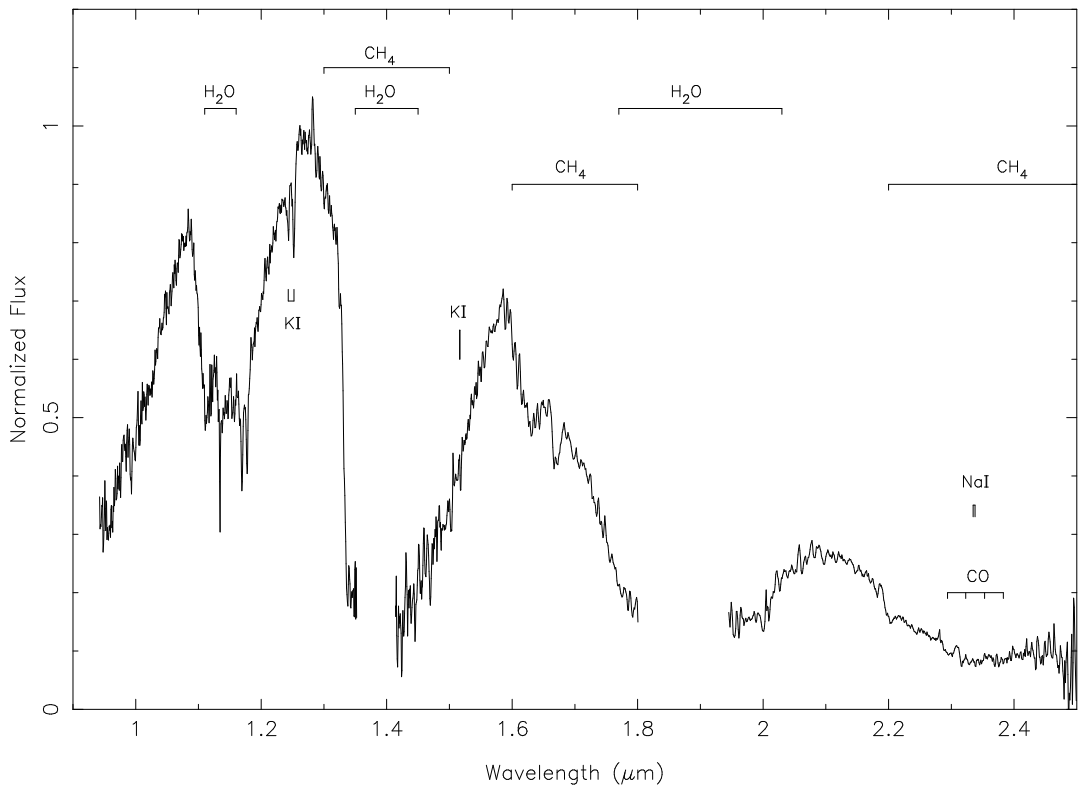}
\caption{ESO NTT SOFI near-infrared spectrum of \thetdwarf{} from 
0.94--2.5\micron. Regions of strong telluric absorption around 1.4 and 
1.9\micron{} have been removed for clarity. The locations of prominent
H$_2$O and CH$_4$ absorption bands in the atmosphere of \thetdwarf{} are 
indicated. Also labelled are the K\,I doublet at 1.25\micron{} (equivalent
widths 3.5 and 7.0$\pm$0.5\AA) and tentative detections of the K\,I doublet 
at 1.52\micron{} and the Na\,I doublet at 2.33\micron. 
}
\label{fig:irspec}
\end{figure*}

\section{Optical and near-infrared photometry}
On the SSS plates, \thetdwarf{} is seen to be very red, clearly visible 
in the $I$ band, but completely invisible at $R$. Examination of the 2MASS 
survey data for this region confirmed \thetdwarf{} to be bright in the 
near-infrared, but with the rather blue near-infrared colours indicative 
of a T~dwarf, rather than an M~or L~dwarf.

The star \thestar{} and its companion \thetdwarf{} lie in a region of the 
2MASS survey for which only the Quicklook survey data have been released
to date, and therefore there are no catalogue magnitudes available. Thus
we used standard aperture photometry techniques to measure the flux from
\thetdwarf{} at $J$, $H$, and $K_s$. The Quicklook data are compressed with a 
lossy algorithm, potentially damaging any derived photometry. 
However, the 2MASS team (Carpenter \cite{carpenter02}) have shown that for 
bright sources such as \thetdwarf, random photometric errors introduced by 
the compression are minimal. To check this, we measured $J$, $H$, and $K_s$
magnitudes from Quicklook images for a sample of 7 M~dwarfs with roughly 
equivalent magnitudes and colours to \thetdwarf, and which also have 
photometric measurements in the 2MASS catalogues from the uncompressed data. 
In this way, we confirmed that the Quicklook photometry is certainly good 
to better than \magnit{0}{1}, adequate for present purposes. Finally, we 
also measured a $K_s$ band magnitude from our SOFI spectroscopy acquisition 
images, yielding a result consistent with the 2MASS value to within 
\magnit{0}{03}. The optical and near-infrared photometry are shown in 
Table~\ref{tab:posphot}.

\begin{table}
\caption[]{Astrometry and photometry for \thetdwarf{} from the SSS and 
2MASS\@. The resulting proper motion is 
$\mu_\alpha\,\cos\delta = +4.131\pm 0.071$ arcsec/yr, 
$\mu_\delta = -2.489\pm 0.025$ arcsec/yr.}
\label{tab:posphot}
\begin{center}
\begin{tabular}{@{}ccrl@{}}
\hline
$\alpha, \delta$ (J2000.0)     & Epoch    & Magnitude   & Data  \\ \hline
$22~04~09.465$~~$-56~46~52.58$ & 1997.771 & $I$=16.59   & SSS   \\
$22~04~10.392$~~$-56~46~57.29$ & 1999.666 & $I$=16.77   & SSS   \\
$22~04~10.517$~~$-56~46~57.78$ & 1999.855 & $J$=12.11   & 2MASS \\
$22~04~10.513$~~$-56~46~57.71$ & 1999.855 & $H$=11.59   & 2MASS \\
$22~04~10.523$~~$-56~46~57.82$ & 1999.855 & $K_s$=11.17 & 2MASS \\
\hline
\end{tabular}
\end{center}
\end{table} 

\section{Near-infrared classification spectroscopy}
We obtained near-infrared (1--2.5\micron) classification spectroscopy for 
\thetdwarf{} using the SOFI camera/spectrometer on the ESO 3.5-m New Technology
Telescope on La Silla on the night of 16--17 November 2002. The conditions 
were photometric and the seeing $\sim$\,1.0 arcsec FWHM\@. The instrument uses 
a 1024$\times$1024 pixel Rockwell HAWAII array with a pixel size of 0.294
arcsec, and uses long slits for spectroscopy. A slit width of 1 arcsec 
was used with the two grisms covering the $J+H$ (blue) and $H+K$ (red) 
spectral regions respectively, at a spectral resolution of 600. Three dithered 
on-chip exposures of 120 seconds were obtained for each grism. Similar
calibration data were obtained for a nearby bright spectral standard 
(CD-57\,8484, spectral type G5).

The data reduction procedure was standard. Each individual spectral image was 
flat-fielded using a tungsten-illuminated spectral dome flat, sky-subtracted
using the mean of the two other dithered images, and a one-dimensional 
spectrum optimally extracted. Wavelength calibration was achieved 
independently for each grism using a xenon arc lamp. Each spectrum was then 
divided by the spectral standard and multiplied back by a G5
template spectrum smoothed to the resolution of each grism. Finally, the 
three individual spectra for each grism were averaged and the two spectral 
regions combined to yield the spectrum seen in Figure~\ref{fig:irspec}.
The total integration time is 360 seconds. 

After reduction, the extracted spectrum was classified in two ways. First,
we made a simple morphological comparison with template T dwarf spectra 
available from Burgasser \etal{} (\cite{burgasser02}), Leggett \etal{} 
(\cite{leggett00}), and Leggett (\cite{leggett02b}). This method yielded 
a classification of T3. Second, we used the spectral classification indices 
for T dwarfs of Geballe \etal{} (\cite{geballe02}) and Burgasser \etal{} 
(\cite{burgasser02}), yielding mean types of T2.75 and T2.25, respectively 
(see Table~\ref{tab:indices}). Consequently, we classify \thetdwarf{} as 
spectral type T2.5, with half a subclass error. 

\begin{table}
\begin{center}
\caption[]{Near-infrared spectral classification indices for \thetdwarf{}
following the schemes of Geballe \etal{} (\cite{geballe02}) and
Burgasser \etal{} (\cite{burgasser02}).} 
\label{tab:indices}
\begin{tabular}{lcl}
\hline
Index                  & Value & Type \\ \hline
\multicolumn{3}{c}{Geballe \etal{} (\cite{geballe02})} \\ \hline
H$_{2}$O~~1.2\micron{} & 1.919 & T2 \\
H$_{2}$O~~1.5\micron{} & 2.600 & T2 \\
CH$_{4}$~~1.6\micron{} & 1.367 & T3 \\
CH$_{4}$~~2.2\micron{} & 1.837 & T2 \\ \hline
\multicolumn{3}{c}{Burgasser \etal{} (\cite{burgasser02})} \\ \hline
H$_{2}$O\_A            & 0.545 & T2--3  \\
H$_{2}$O\_B            & 0.607 & T1--2  \\
CH$_{4}$\_A            & 0.890 & T2--3  \\
CH$_{4}$\_B            & 0.695 & T3     \\
CH$_{4}$\_C            & 0.516 & T2--3  \\
HJ                     & 0.568 & T3     \\
KJ                     & 0.231 & T3     \\
2.11/2.07              & 1.011 & T5     \\ \hline
\end{tabular}
\end{center}
\end{table}

\section{Physical properties}
Given the distance to \thetdwarf, we can determine its absolute near-infrared 
magnitudes as M$_J$=\magnit{14}{31}, M$_H$=\magnit{13}{79}, and 
M$_{K_s}$=\magnit{13}{37}, with errors of less than $\pm$\magnit{0}{1}. 
In the absence of reliable optical and thermal
infrared data, we defer the direct calculation of the bolometric magnitude
of \thetdwarf{} to a later paper. Here, for a first order luminosity estimate,
we must assume a bolometric correction, which is unfortunately not
yet well-determined for early T dwarfs. Reid (\cite{reid02d}) gives a
BC$_{K_s}$=\magnit{3}{3} for the slightly earlier T2 dwarf SDSS\,1254$-$01 
(\cf{} Dahn \etal{} \cite{dahn02}), while an approximate interpolation between 
L8 and T6 in the data of Leggett \etal{} (\cite{leggett02}; their Fig.\ 3) 
yields a BC$_{K_s}$=\magnit{3}{0}--\magnit{3}{1} for 
$J-K$\,$\sim$\,\magnit{0}{9}. Finally, the COND atmospheric models of 
Baraffe \etal{} (personal communication) would suggest a 
BC$_{K_s}$\,$\sim$\,\magnit{2}{8}--\magnit{3}{1} for sources with 
T$_{\rm eff}$\,$\sim$\,1000--1300\,K as appropriate for early 
T dwarfs. Thus here we assume a BC$_{K_s}$=\magnit{3}{0}$\pm$\magnit{0}{1} 
for T2.5, which then yields M$_{\rm bol}$=\magnit{16}{37} and 
log\,L/\Lsolar=$-4.67$ ($\pm$20\%) for \thetdwarf, assuming 
M$_{\rm bol}$=\magnit{4}{69} for the Sun.

Next, using the M$_{\rm bol}$--radius relation derived by Dahn \etal{} 
(\cite{dahn02}) from theoretical models and renormalised by Reid 
(\cite{reid02d})
$$R/\Rsolar = 0.097 - 0.00077 (M_{\rm bol} - 16.2)^{2.9}~,$$
we obtain a radius of 0.097\Rsolar{} for \thetdwarf{} ($\sim$\,1 Jupiter 
radius, as is roughly true for all very late type stars, brown dwarfs, and 
gas giant planets supported by electron degeneracy pressure). Then, using
${\rm L_{bol}}\propto r^2{\rm T_{eff}}$ and adopting $\rm T_{eff}$=5771\,K
for the sun, we calculate an effective temperature of $\sim$1260\,K 
($\pm$\,60\,K) for \thetdwarf, consistent with a spectral type of T2.5 
(Burgasser \etal{} \cite{burgasser02}). 

To estimate the mass of \thetdwarf, we need an estimate of its age, as 
objects below the hydrogen fusion limit continuously cool and grow dimmer 
over their lifetimes. Fortunately, as it is a companion to a well-studied
bright star, we can use age determinations for the latter as a proxy.
Lachaume \etal{} (\cite{lachaume99}) examined several possible age dating
techniques for nearby main sequence stars, and for \thestar{} settled on a 
an age range of 0.8--2\,Gyr, based on its rotational properties.

Using the model isochrones of Burrows \etal{} (\cite{burrows97}), we find
that for a luminosity of log\,L/\Lsolar=$-4.67$, an age range of 0.8--2\,Gyr
corresponds to a mass range of 40--60\Mjup, with the median age of 1.3\,Gyr
quoted by Lachaume \etal{} (\cite{lachaume99}) yielding a mass of 50\Mjup. 
As the mass estimate is essentially unaffected by the errors in our luminosity 
determination for \thetdwarf, it appears to be a genuine brown dwarf, as 
expected. In addition, consistency in the suite of derived parameters is 
confirmed using the 1\,Gyr models of Chabrier \etal{} (\cite{chabrier00}), 
where our luminosity estimate implies R=0.097\Rsolar, ${\rm T_{eff}}$=1270\,K, 
and M\,$\sim$\,43\Mjup. 

Finally, it is worth noting that Endl \etal{} (\cite{endl02}) have monitored 
\thestar{} for radial velocity changes indicative of a planetary companion.
They found no periodic signal in their data, but did draw attention to a 
long-term trend which they suggested might be due to a low-mass stellar or 
brown dwarf companion. Given the parameters we have determined for \thetdwarf, 
it is clear that it is not the source of this long-term velocity change, but
it remains possible that the \thestar{} system is a hierarchical multiple, 
similar to the Gl570\,ABCD system (Burgasser \etal{} \cite{burgasser00}). 
Planned HST direct imaging observations of \thestar{} could have located 
close brown dwarf or giant planet companions, but unfortunately failed due 
to guide star acquisition problems (Schroeder \etal{} \cite{schroeder00}).

\section{Conclusions}
Through a survey for high proper motion objects, we have identified 
\thetdwarf, a T2.5 dwarf, the brightest object in the T spectral class, and 
the nearest bona fide brown dwarf to the Sun. In combination with its 
accurately known distance, relatively well-known age, and large separation
from its primary star, these characteristics make \thetdwarf{} a new benchmark 
in the study of substellar objects, amenable to a wide range of detailed
atmospheric and chemical observations. Urgently needed, however, are more 
accurate near-infrared photometry, including measurements out to the $L$ 
band, and spectral model fitting to derive an explicit bolometric magnitude 
for \thetdwarf, and thus help calibrate the presently poorly-understand 
L-T dwarf boundary. 

\begin{acknowledgements}
We would like to thank Lutz Wisotzki for helping arrange the taking
of near-infrared classification spectra during an unrelated NTT run on 
very short notice. This discovery is based on the excellent data of the 
SuperCOSMOS Sky Surveys at the Wide-Field Astronomy Unit of the Institute 
for Astronomy, University of Edinburgh and we would like to thank Nigel 
Hambly for his advice on the use of SSS data. We also thank the referee
for some helpful comments. We have made use of data 
products from the Two Micron All Sky Survey, a joint project of the 
University of Massachusetts and IPAC, funded by NASA and the NSF, and of 
the VizieR catalogue access tool, CDS, Strasbourg. NL thanks the 
EC Research Training Network ``The Formation and Evolution of Young Stellar 
Clusters'' (HPRN-CT-2000-00155) for financial support.
\end{acknowledgements}

{}

\end{document}